\documentclass[lettersize,journal]{IEEEtran}

\usepackage{amsmath,amsfonts,amssymb}
\usepackage{amsthm} 
\usepackage{algorithm,algorithmic}
\usepackage{array}
\usepackage[caption=false,font=normalsize,labelfont=sf,textfont=sf]{subfig}
\usepackage{textcomp}
\usepackage{stfloats}
\usepackage{url}
\usepackage{verbatim}
\usepackage{graphicx}
\usepackage{diagbox,multirow,makecell}
\newtheorem{theorem}{Theorem}
\usepackage{balance}
\usepackage{subcaption,caption,float}
\usepackage{color}

\makeatletter 
\let\myorg@bibitem\bibitem
\def\bibitem#1#2\par{%
  \@ifundefined{bibitem@#1}{%
    \myorg@bibitem{#1}#2\par
  }{%
    \begingroup
      \color{\csname bibitem@#1\endcsname}%
      \myorg@bibitem{#1}#2\par
    \endgroup
  }%
}
\makeatother 

\begin{document}
\title{Unsupervised Learning for Joint Beamforming Design in RIS-aided ISAC Systems}
\author{Junjie~Ye, \IEEEmembership{Student Member,~IEEE,}  Lei~Huang, \IEEEmembership{Senior Member,~IEEE,} Zhen~Chen, \IEEEmembership{Senior Member,~IEEE,} Peichang~Zhang,    Mohamed~Rihan, \IEEEmembership{Senior Member,~IEEE}
\thanks{J. Ye, L. Huang, Z. Chen, P. Zhang with the State Key Laboratory of Radio Frequency Heterogeneous Integration (Shenzhen University), Shenzhen 518060, China; M. Rihan is with the Department of Communications Engineering, University of Bremen, Bremen 28359, Germany and also with Department of Electronics and Electrical Communication Engineering, Faculty of Electronic Engineering, Menouf 32952, Egypt. (e-mail: 2152432003@email.szu.edu.cn; lhuang@szu.edu.cn; chenz.scut@gmail.com; pzhang@szu.edu.cn; mohamed.elmelegy@el-eng.menofia.edu.eg). (Corresponding author: Lei Huang.)}
\thanks{This work was supported in part by the National Science Fund for Distinguished Young Scholars under Grant 61925108, the Key Project of International Cooperation and Exchanges of the National Natural Science Foundation of China under Grant 62220106009, the project of Shenzhen Peacock Plan Teams under Grant KQTD20210811090051046, Research Team Cultivation Program of Shenzhen University under Grant 2023DFT003 and the 2035 Program of Shenzhen University under Grant 2022B009.}}
\maketitle

\begin{abstract}
It is critical to design efficient beamforming in reconfigurable intelligent surface (RIS)-aided integrated sensing and communication (ISAC) systems for enhancing spectrum utilization. However, conventional methods often have limitations, either incurring high computational complexity due to iterative algorithms or sacrificing performance when using heuristic methods. To simultaneously achieve both low complexity and high spectrum efficiency, lightweight structures are employed to develop an unsupervised learning-based beamforming design in this work. We tailor image-shaped channel samples and develop an ISAC beamforming neural network (IBF-Net) model. By leveraging unsupervised learning, the loss function incorporates key performance metrics like sensing and communication channel correlation and sensing channel gain, eliminating the need for labeling. Simulations show that the proposed method achieves competitive performance compared to the benchmarks and significantly reduces the computational complexity.
\end{abstract}

\begin{IEEEkeywords}
ISAC, RIS, beamforming design, lightweight network, unsupervised learning.
\end{IEEEkeywords}

\vspace{-0.2cm}
\section{Introduction}
\IEEEPARstart{R}{econfigurable} intelligent surface (RIS) and integrated sensing and communication (ISAC) are two promising technologies that empower future communication networks \cite{RISISAC_survey2}. ISAC facilitates the coexistence of sensing and communication systems within the same spectrum, while RIS enables channel adjustments to enhance spectrum efficiency \cite{chen2} \cite{czz}. Given their potential, numerous research endeavors have explored RIS-aided ISAC systems. In \cite{RIS_ISAC_jiang}, the authors employed semi-definite relaxation (SDR) to maximize radar signal-to-noise ratio (SNR) iteratively. Similarly, a successive lower-bound maximization approach was proposed for beamforming design in a comparable scenario in \cite{Zhongxin}. Moreover, \cite{RIS_ISAC_liu} considered optimizing radar signal-to-interference-plus-noise ratio (SINR) under various communication constraints. Despite exhibiting superior performance, these methods entail high computational complexities due to iterations. In contrast, a heuristic non-iterative method was introduced in \cite{RIS_ISAC_fan}, which designed RIS phases through an innovative subspace rotation method. Albeit the non-iterative method is time efficient, its solution is obtained via direct gradient descent, rendering it susceptible to sub-optimality.

Concurrently, deep learning (DL) has attracted great attentions for efficiently addressing highly non-convex problems. Therefore, numerous studies explored DL-aided beamforming designs. In \cite{RIS_supervised}, a fully connected network (FCN) was utilized to design RIS phases through supervised learning. Conversely, in \cite{RIS_Un} and \cite{RIS_Un_transfer}, unsupervised learning was employed to train RIS beamforming networks, thereby eliminating extensive labeling process. To simplify the model, a lightweight network was proposed for RIS phase design in \cite{My}. Reinforcement learning (RL) was employed to optimize RIS phases in \cite{RL_MISO}. Beyond RIS phases design, DL also aids in ISAC beamforming. In \cite{ISAC_predictbf}, a FCN was trained to predict tracking beams, while \cite{ISAC_Un_bf} introduced unsupervised learning to manage interference in ISAC systems. In \cite{ISAC_RL}, RL was leveraged to allocate resources in ISAC systems. However, these endeavors primarily focused on either RIS or ISAC beamformer design.

In this paper, we propose an unsupervised lightweight DL approach for beamforming design in a RIS-aided ISAC system, aiming to simultaneously reduce complexities and ensure adequate performance. The image-shaped channel samples are elaborately designed for channel feature extractions and a lightweight ISAC beamforming neural network (IBF-Net) is developed by leveraging the simple and effective lightweight components. To mitigate labeling overhead, we devise a loss function based on unsupervised learning, considering the equilibrium between sensing and communication channel correlation, along with sensing channel gain. Through simulations, we validate the efficacy of our approach, demonstrating its ability to expedite beamforming design and reduce performance degradation. The code is available online (https://github.com/Yejacky456/DL-Beamforming-RIS-ISAC).

\vspace{-0.3cm}
\section{System Model And Formulations} \label{Sec_model}
We consider a RIS-aided ISAC system, where a $M$-antenna ISAC base station (BS) is deployed to simultaneously sense a target and serve a single-antenna user. The RIS has $N$ passive elements to adjust the channel environment. Specifically, the BS transmits ISAC signal $\mathbf{w}s$, where $s$ is the unit-power baseband signal and $\mathbf{w} \!\!\in \!\!\mathbb{C}^{M\times1}$ denotes the transmit beamformer. The ISAC signal propagates through the channel $\mathbf{G}\! \in \!\!\mathbb{C}^{N\times M}$ from BS to RIS. After tuning phases at the RIS, the signal propagates through the channel $\mathbf{h}_{r,c}^H \!\in\! \mathbb{C}^{1 \times N}$ to the user while illuminates the target through the channel $\mathbf{h}_{r,t}^H \!\in\! \mathbb{C}^{1 \times N}$. With the target illuminated, the echo is reflected at the RIS and received by the BS. Consequently, the signal received by the user is given as
\begin{align} \label{com_model}
    y_c &= \mathbf{h}_{r,c}^H \mathbf{\Theta}^H \mathbf{G} \mathbf{w }s + n_c,
\end{align}
in which $\mathbf{\Theta} \!= \!\mathrm{diag} \left( \!e^{j\widetilde{\theta}_1}, e^{j\widetilde{\theta}_2},\dots,e^{j\widetilde{\theta}_N} \!\right)\! \in \!\mathbb{C}^{N \times N}$ represents the matrix of phase shifts, wherein the diagonal elements correspond to the phase shifts of each RIS element. The term $n_c$ is the noise at the user, assumed to adhere to additive white Gaussian noise (AWGN) with a distribution of $\mathcal{CN}(0,\sigma_c^2)$. Utilizing the reception model delineated in (\ref{com_model}), the SNR at the user can be mathematically expressed as
\begin{align}
    \gamma_c &= \frac{|\mathbf{h}_c^H \mathbf{w}|^2}{ \sigma_c^2},
\end{align}
where $\mathbf{h}_c^H \!\!=\!\! \mathbf{h}_{r,c}^H \mathbf{\Theta}^H \mathbf{G}$ represents the end-to-end communication channel. Alternatively, $\mathbf{h}_c^H$ is also given as $\mathbf{h}_c^H \!\!=\!\! \mathbf{v}^H \overline{\mathbf{\Phi}}_c$, in which $\overline{\mathbf{\Phi}}_c \!\!=\!\! \mathrm{diag}(\mathbf{h}_{r,c}^H)\mathbf{G} \!\in \!\mathbb{C}^{N \times M}$ denotes the cascaded channel towards the user, and $\mathbf{v}\!\!=\!\!\mathrm{diag}(\mathbf{\Theta})\!\in\! \mathbb{C}^{N \times 1}$.

Similarly, the target echo received by the BS is given as
\begin{align} \label{rad_model}
    \mathbf{y}_r &= \mathbf{G}^H \mathbf{\Theta}\mathbf{h}_{r,t}\mathbf{h}_{r,t}^H \mathbf{\Theta}^H\mathbf{G} \mathbf{w}s + \mathbf{n}_r,
\end{align}
where $\mathbf{n}_r \!\in\! \mathbb{C}^{N \times 1}$ denotes the AWGN at BS with a distribution of $\mathcal{CN}(0,\sigma_r^2 \mathbf{I}_M)$.
The echo SNR is then given as
\begin{align}
    \gamma_r &= \frac{\|\mathbf{H}_t \mathbf{w}\|^2}{\sigma_r^2},
\end{align}
wherein $\mathbf{H}_t = \mathbf{h}_t\mathbf{h}_t^H$ and $ \mathbf{h}_t^H = \mathbf{h}_{r,t}^H \mathbf{\Theta}^H \mathbf{G}$. Also, $\mathbf{h}_t^H$ can be reformulated as $\mathbf{v}^H \overline{\mathbf{\Phi}}_t$, where $\overline{\mathbf{\Phi}}_t = \mathrm{diag}(\mathbf{h}_{r,t}^H)\mathbf{G} \in \mathbb{C}^{N \times M}$.

In this scenario, we aim to maximize the sensing SNR of the echo signal by jointly optimizing transmit beamformer $\mathbf{w}$ and RIS phase shift matrix $\mathbf{\Theta}$. The corresponding optimization problem can be formulated as
\begin{subequations}\label{formulation}
\begin{align} 
\vspace{-1cm}
\max _{\mathbf{w},\mathbf{\Theta}} &~  \gamma_r \tag{\ref{formulation}{a}} \\ 
s.t.&~  \gamma_c \geq \tau_c, \tag{\ref{formulation}{b}} \\
&~ \|\mathbf{w}\|^2 \leq P_{t}, \tag{\ref{formulation}{c}} \\
&~ |\mathbf{\Theta}_{n,n}| = 1, \forall n=1,\dots,N. \tag{\ref{formulation}{d}}
\end{align}
\end{subequations}
The constraint (\ref{formulation}{b}) imposes a lower bound threshold $\tau_c$ on the user SNR, while the constraint (\ref{formulation}{c}) delineates the transmit power budget. Additionally, the constraint (\ref{formulation}{d}) enforces the unit-modulus nature of the RIS. Various methods have been proposed to address the problem (\ref{formulation}). However, these approaches often have elevated complexities stemming from iterative procedures or performance degradation due to sub-optimal solutions.

\section{Proposed Unsupervised Learning Approach}
In pursuit of simultaneously mitigating computational complexities and minimizing performance degradations, we introduce an unsupervised lightweight learning scheme for beamforming design in this section. Specifically, a closed-form expression for the transmit beamformer is provided first. Subsequently, we delve into the design of the unsupervised learning scheme for optimizing the RIS beamformer.
\subsection{Transmit Beamforming Design} \label{sec_trans_bf}
When RIS beamformer $\mathbf{\Theta}$ is determined, channels $\mathbf{H}_t$ and $\mathbf{h}_c^H$ are rendered constant. Consequently, the task becomes the optimization of the transmit beamformer, given as
\begin{subequations}\label{formulation_w}
\begin{align} 
\max _{\mathbf{w}} &~ \|\mathbf{H}_t \mathbf{w}\|^2 \tag{\ref{formulation_w}{a}} \\ 
s.t.&~  |\mathbf{h}_c^H \mathbf{w}|^2 \geq \tau_c \sigma_c^2, ~~\|\mathbf{w}\|^2 \leq P_{t}.\tag{\ref{formulation_w}{b}}
\end{align}
\end{subequations}
The solution to problem (\ref{formulation_w}) is given in the following theorem:
\begin{theorem}
The optimal transmit beamformer $\mathbf{w}$ is 
\begin{equation} \label{optimize_w}
   \mathbf{w} = \begin{cases} 
   \sqrt{P_t} \frac{\mathbf{h}_t}{\|\mathbf{h}_t\|}, & \text{if }  P_{t} |\mathbf{h}_c^H \mathbf{h}_t |^2\geq \tau_c \sigma_c^2
 \|\mathbf{h}_t\|^2,\\
   x_1 \mathbf{u}_1 + x_2 \mathbf{u}_2, & \text{otherwise},
   \end{cases}
\end{equation}
where
\begin{align}
    \mathbf{u}_1 = \frac{\mathbf{h}_c}{\|\mathbf{h}_c\|},  \quad \mathbf{u}_2 = \frac{\mathbf{h}_t-(\mathbf{u}_1^H \mathbf{h}_t)\mathbf{u}_1}{\|\mathbf{h}_t-(\mathbf{u}_1^H \mathbf{h}_t)\mathbf{u}_1\|},
\end{align}
and
\begin{align}
    \mathbf{x}_1 \!=\! \sqrt{\frac{\tau_c \sigma_c^2}{\|\mathbf{h}_c\|^2}}\frac{\mathbf{u}_1^H \mathbf{h}_t}{|\mathbf{u}_1^H \mathbf{h}_t|}, \!\! \quad \!\!\mathbf{x}_2 \!= \!\sqrt{P_{t}\!-\!\frac{\tau_c \sigma_c^2}{\|\mathbf{h}_c\|^2}}\frac{\mathbf{u}_2^H \mathbf{h}_t}{|\mathbf{u}_2^H \mathbf{h}_t|}.
\end{align}
\label{theor-1}
\end{theorem}
\begin{proof}
The detailed derivations can be referred to \cite{closeform}. 
\end{proof}

Drawing from \textbf{Theorem \ref{theor-1}}, a strong correlation between the two channels, given as $|\mathbf{h}_c^H \mathbf{h}_t |^2$, leads to a substantial overlap between the sensing and communication channel sub-spaces. Hence, even if the transmit beamformer aligns with sensing channel, power can be effectively reused for communication, thus directly fulfilling the user SNR requirement. Conversely, when the channel correlation is weak, the direct alignment beamformer falls short of satisfying the user SNR constraint. In such cases, the beamformer needs to be situated within the expanded subspace of $\mathbf{h}_t$ and $\mathbf{h}_c$ to maximize the echo SNR while concurrently meeting the user SNR requirement.

\vspace{-0.1cm}
\subsection{RIS Beamforming Design}
\subsubsection{Sample Constructions} \label{sample}
Our primary objective is to enhance the gains of the channels $\mathbf{H}_t$ and $\mathbf{h}_c$. To achieve this, the samples are structured by extracting information from $\mathbf{H}_t$ and $\mathbf{h}_c$. As discussed in Sec.\ref{Sec_model}, $\mathbf{H}_t$ and $\mathbf{h}_c$ can be expressed as $\mathbf{H}_t\!\!=\!\!\overline{\mathbf{\Phi}}_t^H\mathbf{v} \mathbf{v}^H \overline{\mathbf{\Phi}}_t$ and $\mathbf{h}_c^H\!\!=\!\!\mathbf{v}^H \overline{\mathbf{\Phi}}_c$. Consequently, the channel gains can be formulated as
\begin{subequations} \label{chan_gain}
    \begin{align}
        \|\mathbf{H}_t\|^2 &=  \mathbf{v}^H \overline{\mathbf{\Phi}}_t  \overline{\mathbf{\Phi}}_t^H \mathbf{v}\mathbf{v}^H \overline{\mathbf{\Phi}}_t  \overline{\mathbf{\Phi}}_t^H \mathbf{v}, \\
        \|\mathbf{h}_c\|^2 &= \mathbf{v}^H \overline{\mathbf{\Phi}}_c  \overline{\mathbf{\Phi}}_c^H \mathbf{v}.
    \end{align}
\end{subequations}
Equations in (\ref{chan_gain}) illustrate that the channel characteristics are encapsulated within $\widetilde{\mathbf{\Phi}}_t \!=\! \overline{\mathbf{\Phi}}_t  \overline{\mathbf{\Phi}}_t^H \!\in \!\mathbb{C}^{N \times N}$ and $\widetilde{\mathbf{\Phi}}_c \!=\! \overline{\mathbf{\Phi}}_c  \overline{\mathbf{\Phi}}_c^H \!\in\! \mathbb{C}^{N \times N}$, separated from the optimization variable $\mathbf{v}$. Therefore, we opt to utilize $\widetilde{\mathbf{\Phi}}_t$ and $\widetilde{\mathbf{\Phi}}_c$ for constructing the samples.

However, conventional neural networks are designed to process real numbers, whereas $\widetilde{\mathbf{\Phi}}_t$ and $\widetilde{\mathbf{\Phi}}_c$ are typically complex. Therefore, it is imperative to transform them into real representations. A straightforward approach is to extract the real and imaginary components, denoted as $\boldsymbol{\mathcal{R}}_t \!=\! \mathfrak{Re}\lbrace\widetilde{\mathbf{\Phi}}_t\rbrace$, $\boldsymbol{\mathcal{I}}_t \!= \!\mathfrak{Im}\lbrace\widetilde{\mathbf{\Phi}}_t\rbrace$, $\boldsymbol{\mathcal{R}}_c \!= \!\mathfrak{Re}\lbrace\widetilde{\mathbf{\Phi}}_c\rbrace$, and $\boldsymbol{\mathcal{I}}_c \!=\! \mathfrak{Im}\lbrace\widetilde{\mathbf{\Phi}}_c\rbrace$, respectively. Subsequently, we expand the dimensions of $\boldsymbol{\mathcal{R}}_t$, $\boldsymbol{\mathcal{I}}_t$, $\boldsymbol{\mathcal{R}}_c$ and $\boldsymbol{\mathcal{I}}_c$ to $\mathbb{R}^{1 \times N \times N}$ and concatenate them along the expanded dimension. Consequently, a constructed sample becomes a 3D tensor with dimensions of $\mathbb{R}^{4 \times N \times N}$.

\subsubsection{Lightweight Network Architecture}
\begin{figure*}[!tp]
        \vspace{-0.5cm}
	\centering
	\includegraphics[scale=0.49]{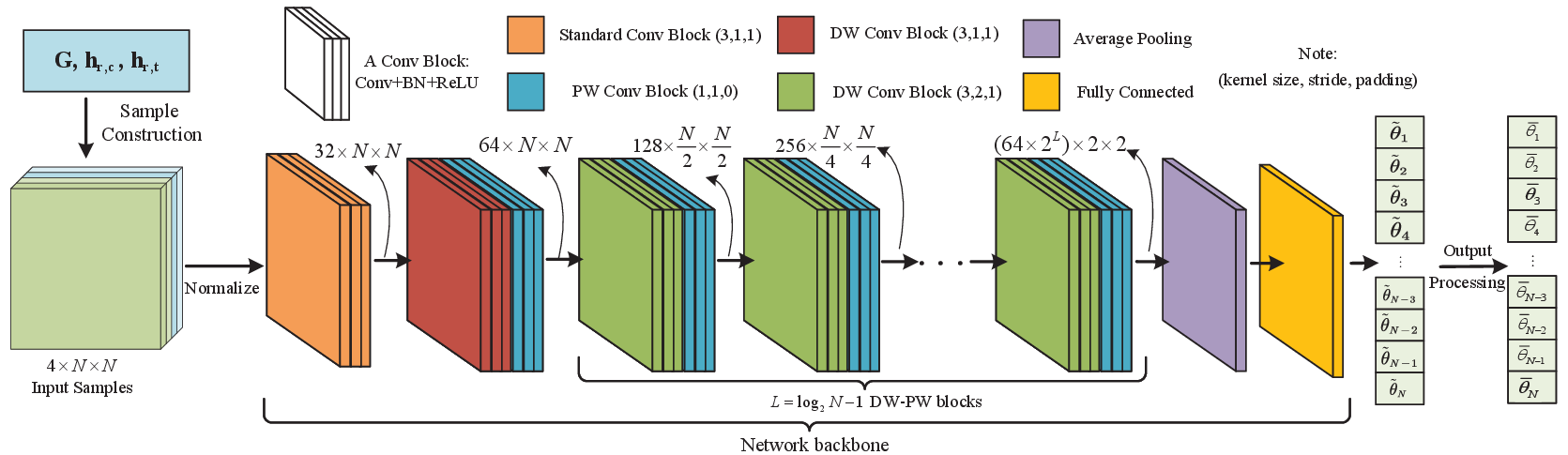}%
        \caption{Network Architecture of IBF-Net.}
	\label{network}
        \vspace{-0.5cm}
\end{figure*}
Considering that excellent system performance and low computational complexities are targeted, we are motivated to design IBF-Net with lightweight structures, which are renowned for their effectiveness and simplicity \cite{mobilenet}. As shown in Fig. \ref{network}, the network comprises convolution blocks (Conv Blocks), an average pooling layer (Avg Pool) and a fully connected layer (FC), where each Conv Block comprises a Conv layer, a batch normalization layer (BN) and a rectified linear unit function (ReLU). The Conv Blocks include one standard Conv Block as well as $L\!+\!1$ lightweight Conv Block, where $L\!\!=\!\!\log_2 N \!\!-\!\!1$. The lightweight Conv Blocks integrate the lightweight structures of depthwise (DW) and pointwise (PW) convolution, thus also termed DW-PW Conv Blocks.  In the standard Conv Block, the kernel size, stride, and padding are set as (3,1,1). For the DW-PW Conv Blocks, DWs have kernel and padding sizes of 3 and 1, while those of PWs are set as 1 and 0. The stride of the first DW is 1 and the remaining $L$ DWs have stride sizes of 2.

Ahead of IBF-Net processing, a normalization operation is performed on the input sample. The standard Conv Block then processes the normalized samples to produce a feature map of size $32\!\!\times\!\! N\!\!\times\!\! N$. This feature map passes through $L\!\!+\!\!1$ DW-PW Blocks, where the map size remains unchanged in the first DW-PW Block while in each subsequent $L$ DW-PW Conv Block, the size is halved and the depth is doubled. Thus, the output feature map has a dimension of $(64\!\times \!2^L) \!\times 2\! \times\! 2$. Finally, the Avg Pool of size 2 and the FC are used to predict $\widetilde{\boldsymbol{\theta}}$.


However, the network output $\widetilde{\boldsymbol{\theta}}$ does not inherently satisfy the constraint (\ref{formulation}{d}). Therefore, post-processing is necessary to ensure adherence to the unit-modulus constraint. Specifically, we apply the Euler formula \cite{My} to $\widetilde{\boldsymbol{\theta}}$, resulting in the expression of
\vspace{-0.1cm}
\begin{equation}
    \overline{\boldsymbol{\theta}} = \cos \widetilde{\boldsymbol{\theta}}+j\sin \widetilde{\boldsymbol{\theta}}.
\end{equation}
Consequently, following this post-processing step, the final output $\overline{\boldsymbol{\theta}}$ represents the designed RIS phases, which can satisfy the constraint (\ref{formulation}{d}).

\subsubsection{Loss Design}
In the training phase, it is crucial to design an effective loss function to facilitate the update of network parameters. Traditional optimization methods for RIS beamformer design through manual labeling are notably time-consuming. To avoid extensive labeling efforts, an unsupervised learning mechanism is employed to formulate the loss function, thereby promoting training for enhancing both sensing and communication channel gains.

Inspired by \cite{RIS_ISAC_fan}, RIS introduces additional channels, effectively expanding the channel subspaces. By adjusting phase shifts, RIS can manipulate the orientation of the subspaces, thereby enhancing the correlations. As elucidated in Sec.\ref{sec_trans_bf}, the increases of channel correlations facilitates power reuse, which enhances the channel gains. Hence, the network is trained to manipulate RIS phase shifts to increase the correlation of the two channels. This gives a loss function as
\vspace{-0.1cm}
\begin{align}
    l_1 = - \frac{1}{S} \sum_{s=1}^S \|\mathbf{H}_t^s \mathbf{h}_c^s\|_2,
\end{align}
where $S$ is the mini-batch size (mbs) during the training phase. As  $l_1$ decreases, an increase of channel correlations occurs, leading to concurrent enhancement of both channel gains.

However, relying solely on $l_1$ as a loss function may yield results that the communication channel gain significantly surpasses the sensing channel gain, as will be demonstrated in Sec.\ref{simulations}. This discrepancy arises from the fact that the sensing channel is subject to double fading in bidirectional propagation, while the communication channel experiences fading only once. Consequently, unless the user SNR threshold is sufficiently high, the communication SNR may substantially exceed the preset threshold, while the radar SNR suffers from performance degradation compared to numerical optimization outcomes. To address this challenge, we introduce the sensing channel gain in the loss function, aiming to balance the correlation of the two channels and the sensing channel gain. This is mathematically expressed as
\vspace{-0.1cm}
\begin{align}
    l_2 = - \frac{1}{S} \sum_{s=1}^S (\|\mathbf{H}_t^s \mathbf{h}_c^s\|_2 + \alpha \|\mathbf{H}_t^s\|_{\mathrm{F}}),
\end{align}
\noindent where $\alpha$ represents a balancing coefficient employed to tradeoff the sensing channel gain and the correlation between the two channels. Different $\alpha$ leads to different training results, where a well-selected $\alpha$ balances the sensing and communication performance while excessively small or large $\alpha$ results in performance bias thereby reducing performance. The effect of varying $\alpha$ will be demonstrated in Sec.\ref{simulations}. Specifically, $l_1$ corresponds to a special case of $l_2$ when $\alpha$ is set to 0.

\vspace{-0.2cm}
\subsection{Overall Algorithm}
In the training phase, an extensive set of channel measurements are gathered to construct samples, as shown in Sec.\ref{sample}. Then, IBF-Net is trained with $l_2$ loss function. In the testing phase, samples are constructed and input to the well-trained network. After obtaining the RIS beamformer from the network, $\mathbf{h}_t$ and $\mathbf{h}_c$ are computed. Subsequently, the transmit beamforming is obtained by using \textbf{Theorem \ref{theor-1}}.

\vspace{-0.1cm}
\section{Simulation Results}\label{simulations}
\begin{figure*}[htbp]
	\centering
	\begin{minipage}{0.32\linewidth}
		\centering
		\includegraphics[width=1.1\linewidth]{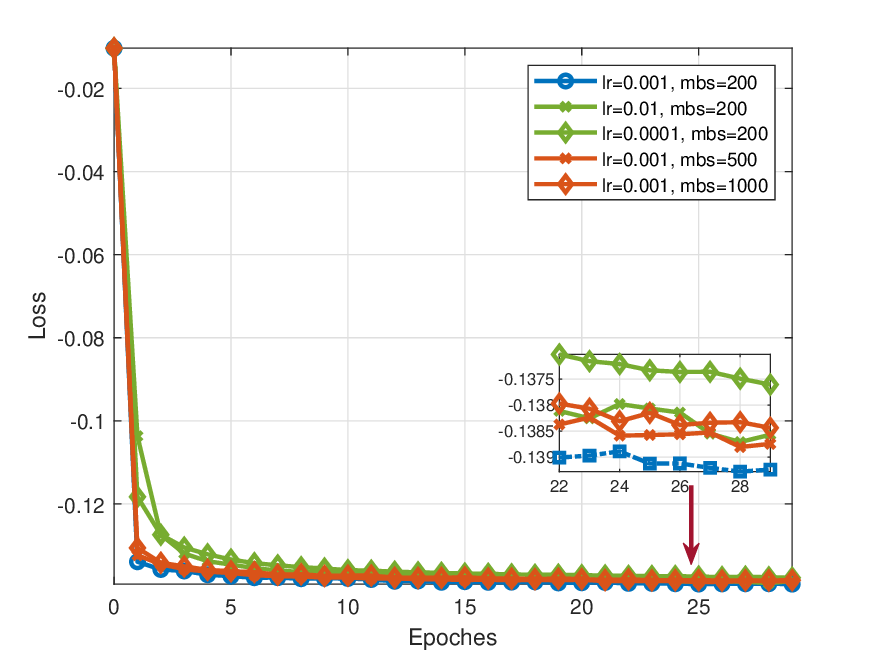} \caption{Changes of losses during training.}
		\label{loss}
	\end{minipage}
	\begin{minipage}{0.32\linewidth}
		\centering		\includegraphics[width=1.1\linewidth]{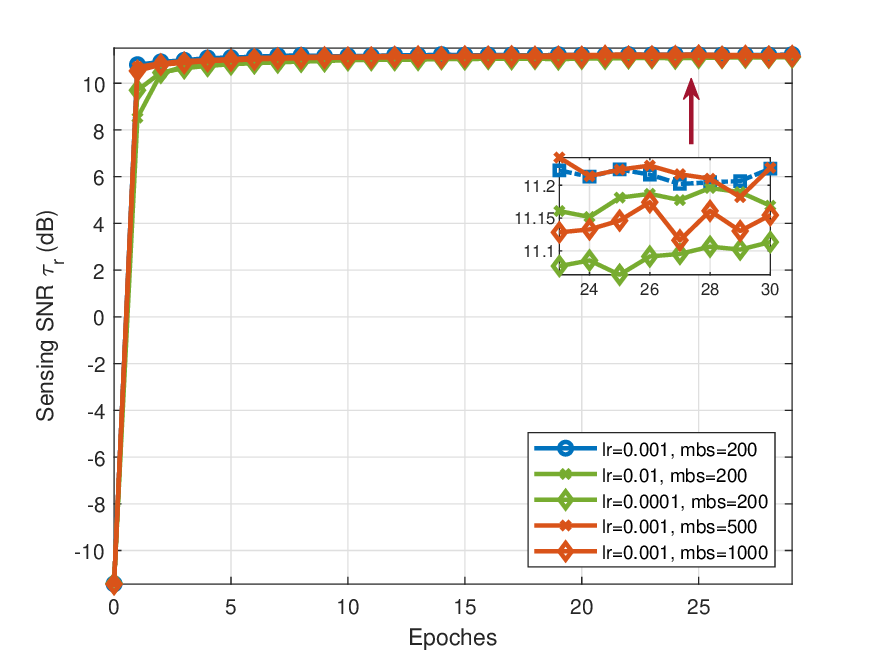}\\
  \caption{Changes of $\tau_r$ during training.}  
		\label{convg}
	\end{minipage}
	\begin{minipage}{0.32\linewidth}
		\centering		\includegraphics[width=1.1\linewidth]{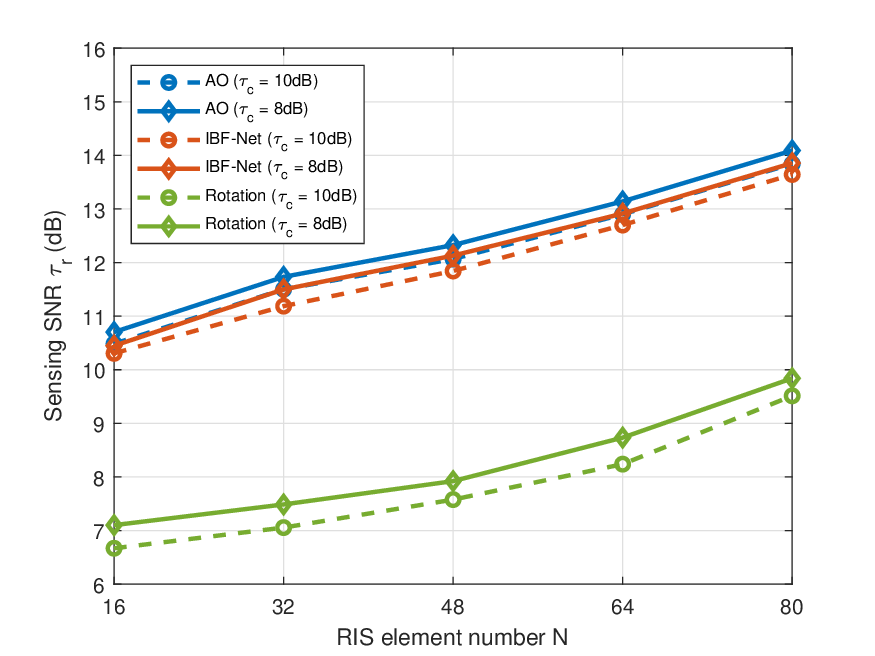}\\
        \caption{RIS element number effect on $\tau_r$.} 
		\label{N}
	\end{minipage}
\end{figure*}

\subsection{Simulation Settings}
Unless stated otherwise, the BS is equipped with $8$ antennas, and the RIS comprises $32$ elements. Additionally, the transmit power is restricted to $P \!\!=\!\! 8$ dBm, while the noise power is set to $\sigma_r^2 \!\!=\!\! \sigma_c^2 \!\!=\!\! -20$ dBm. The user SNR threshold $\tau_c$ is established at $10$ dB. All channels are characterized as Rician channels similar to \cite{My} and assumed to remain stationary random within a long-time range.
For training purposes, the IBF-Net is implemented with PyTorch framework, trained on an NVIDIA Tesla P40 GPU. We generate 500,000 samples to train the model, with training mbs and learning rate (lr) set to 200 and 0.001, unless specific specifications. Additionally, the balancing coefficient $\alpha$ is set to 0.8. During the training process, we utilize the Adam optimizer to update network parameters for 30 epochs. For testing purposes, 100 samples are used to evaluate the generalization capability of the model.

To evaluate the proposed method, we compare it against two representative algorithms. The first benchmark, described in \cite{RIS_ISAC_jiang}, alternately optimizes active and passive beamformers with SDR, yielding near-optimal solutions. Conversely, the subspace rotation approach proposed in \cite{RIS_ISAC_fan} serves as the second baseline, featuring rapid designs for RIS and transmit beamformers through non-alternative dual-variable optimization. For ease of expression, we denote the first benchmark as ``AO'' and abbreviate the second baseline as ``Rotation''.

\vspace{-0.2cm}
\subsection{Performance Evaluations}
We first show the evolution of loss during the training phase in Fig. \ref{loss}. The network undergoes training across various combinations of lr and mbs, where the lr ranges from 0.01 to 0.0001, and mbs is set to 200, 500, and 1000, respectively. Albeit with slight discrepancies in the convergence trajectories, all settings exhibit fast convergence due to proper optimizer selection, careful parameter setting, and stationary sample distribution.  Notably, cases with $\mathrm{lr}\!=\!\!0.001$ demonstrate faster initial decreases in loss, converging to lower levels compared to other cases. Conversely, scenarios with $\mathrm{lr}\!=\!0.01$ initially converge slowly but eventually surpass those with $\mathrm{lr}\!=\!0.0001$, owing to their larger step sizes. Meanwhile, cases with $\mathrm{lr}\!\!=\!\!0.0001$ continue reducing in loss beyond 30 epochs. Furthermore, observations across different mbs reveal a marginal enhancement in convergence with smaller mbs.

The variations in the sensing SNR ($\tau_r$) of the testing samples during  training phase are shown in Fig. \ref{convg}. The results are obtained under the same settings as the previous experiment. Evidently, the average sensing SNRs show a consistent increase throughout the training process across all cases, affirming the efficacy of the proposed method. Moreover, by comparing the $\tau_r$ under different settings, it is notable that the evolution trend of $\tau_r$ is closely related to the changes of loss.

Fig. \ref{N} shows the impact of RIS element number on sensing SNR, alongside comparisons with benchmarks. As the element number ($N$) increases, the sensing SNRs of all approaches rise, which aligns with intuitive expectations as more elements provide greater degrees of freedom. Furthermore, the proposed scheme demonstrates significantly closer performance to the ``AO'' scheme than the ``Rotation'' scheme. This is attributed to the neural network's ability to handle highly non-convex problems, facilitating avoidance from local optima. Additionally, an inverse relationship is observed between the sensing SNR ($\gamma_r$) and the user SNR threshold ($\tau_c$) as $\tau_c$ increases. This phenomenon underscores the spectrum resource tradeoff between sensing and communication.

To assess the impact of the balancing coefficient $\alpha$, IBF-Net is trained using varying $\alpha$ under different configurations, and the results are shown in Table \ref{penalty}. Observations reveal that as $\alpha$ ranges from 0 to 1.6, the sensing SNR ($\gamma_r$) gradually increases while the communication SNR ($\gamma_c$) decreases, getting closer to the user SNR threshold ($\tau_c$). This is because $\alpha$ effectively balances the gains of the sensing channel and the correlation between the channels to facilitate resource allocation. Specifically, when $\alpha\!\!=\!\!0$ (i.e. $l_1$ is considered), the $\gamma_c$ significantly exceeds the preset threshold $\tau_c$ at the cost of reducing $\gamma_r$. As $\alpha$ increases, $\gamma_r$ experiences slight degradation while $\gamma_c$ progressively fails to meet the constraint. With the increase of $\alpha$, the design of $\mathbf{\Theta}$ becomes biased to sensing but worsens the communication channel conditions. To avoid failures of the communication constraint, communication is favored by $\mathbf{w}$. Consequently, the effects of $\mathbf{\Theta}$ and $\mathbf{w}$ on sensing and on communication counteract, leading to $\gamma_r$ degradation and progressive failure of the constraint. Conversely, when $\alpha$ is less than zero, both $\gamma_r$ and $\gamma_c$ exhibit sharp declines, as negative $\alpha$ adversely affects the sensing channel, diminishing the correlation of the channels. Furthermore, as transmit power increases or user constraints relax, the permissible range of $\alpha$ widens. This implies the importance of selecting $\alpha$, as excessively large or small values yield unfavorable outcomes.

\begin{table*}[!htbp]
\caption{Sensing SNR $\gamma_r$ and communication SNR $\gamma_c$ under different balancing factor $\alpha$}
\label{penalty}
\centering
\scalebox{0.92}{
\begin{tabular}{|c|c|c|c|c|c|c|c|c|c|}
\hline
\multirow{2}*{Different Settings} & $\gamma_r$ \& $\gamma_c$  & \multicolumn{8}{c|}{Varying $\alpha$} \\
\cline{3-10}
& (dB) & -0.8& -0.4 & 0 & 0.4& 0.8 & 1.6 & 3.2 & 4.8 \\
\hline
\multirow{2}*{$\tau_c$ = 10dB, P = 8dBm} &  sensing SNR $\gamma_r$  & -19.96 &  -0.25  &  8.72  & 10.74  & 11.22 &  11.3  & 10.93  & 10.77\\
\cline{2-10}
  & Comm. SNR $\gamma_c$   & 9.89  & 11.69 &  17.38 &  12.82 &  10.59 &  10.18  & 10.01  &  9.98  \\
\hline 
\multirow{2}*{$\tau_c$ = 8dB, P = 8dBm} &  sensing SNR $\gamma_r$   & -21.19  & -0.16  &  8.90  & 10.89 &  11.31  & 11.52  & 11.45  & 11.37 \\
\cline{2-10}
  & Comm. SNR $\gamma_c$   & 8.08  & 21.10  & 17.25 &  13.58 &  10.71  &  9.08  &  8.46  &  8.26  \\
\hline 
\multirow{2}*{$\tau_c$ = 8dB, P = 10dBm} & sensing SNR $\gamma_r$   & -19.55 &   1.54  & 10.76  & 12.77  & 13.27 & 13.48  & 13.41  & 13.37\\
\cline{2-10}
  & Comm. SNR $\gamma_c$  & 10  & 23.07 &  19.4 &  15.11  & 11.95 &  10.5  & 10.34  & 10.11  \\
\hline 
\end{tabular}}
\end{table*}

\vspace{-0.3cm}
\subsection{Complexity Comparisons}
We compare the computational complexities of the proposed network and the benchmarks in this subsection. According to \cite{mobilenet}, the additions and multiplications operation number of the proposed network can be given as $1184 N^2 \!+\! 576 (N^2\!-\!N) \!+\! 2048  N^2\log_2N  \!+\!  6 M$, where its complexity can be given as $\mathcal{O}\left(N^2  + N^2\log_2N  +  M\right)$. For the benchmarks, the complexities are given as $\mathcal{O}\left(I_0 \left(M^{4.5}\log(1/\epsilon)\!+\!M^{3}\!+\!N^{4.5}\log(1/\epsilon)\!+\!N^{3}\right)\right)$ for ``AO'' and $ \mathcal{O}\left(I_1 \left((M\!+\!1)N^2\! +\! N^2 \!+\! M N\!+\!M\!+\!N\right) \!+\! M\right)$  for ``Rotation'', where $I_0$ and $I_1$ indicates the iteration numbers and $\epsilon$ is the algorithm accuracy. In addition, the running times of the proposed method and the benchmarks are shown in Table \ref{running_time}. We compute the running time by using the average durations for designing the system beamformers with $100$ channel realizations. The results in Table \ref{running_time} show that the proposed method consumes the least time compared to the benchmarks. Specifically, the ``AO'' approach consumes nearly $1000$ times more computational resources than the proposed method. Additionally, the time complexity of the ``Rotation" scheme is marginally higher than that of the proposed method. This difference arises from the gradient descent utilized by the ``Rotation'' scheme, which entails an iterative gradient search process. In summary, the proposed method achieves high performance and significantly reduces computational resources, thus demonstrating its efficiency and effectiveness.
\vspace{-0.2cm}
\begin{table}[!ht]
\caption{Average running time of one sample (ms)}
\label{running_time}
\centering
\scalebox{0.88}{
\begin{tabular}{|c|c|c|c|}
\hline
\diagbox{Parameters}{Time}{Algorithm} & AO & IBF-Net & Rotation \\
\hline
N = 16 & $3.63\times10^3$ & 3.7 & 4.2 \\
\hline
N = 32 & $6.61\times10^3$ & 3.9 & 5.4  \\ 
\hline
N = 48 & $13.65\times10^3$ & 4.1 &  11.0 \\ 
\hline
N = 64 & $22.71\times10^3$  & 4.6 &  16.3 \\ 
\hline
N = 80 & $37.76\times10^3$ & 5.2 & 20.2 \\
\hline
\end{tabular}}
\end{table}

\vspace{-0.5cm}
\section{Conclusions}
This paper introduced an unsupervised learning method for beamforming design in a RIS-aided ISAC system. Specifically, a lightweight IBF-Net model was developed for simple and effective beamforming design, leveraging the customized image-shaped channel samples. Moreover, we formulated a loss function to balance the sensing and communication channel correlations, as well as the sensing channel gain. Subsequently, the transmit beamformer was obtained by a closed-form expression. Simulations verified that our proposed unsupervised learning-based beamforming method yielded satisfying performance and substantially reduced computational complexities.
\vspace{-0.5cm}
\bibliographystyle{IEEEtran}
\bibliography{literature}{}

\begin{thebibliography}{10}
\providecommand{\url}[1]{#1}
\csname url@samestyle\endcsname
\providecommand{\newblock}{\relax}
\providecommand{\bibinfo}[2]{#2}
\providecommand{\BIBentrySTDinterwordspacing}{\spaceskip=0pt\relax}
\providecommand{\BIBentryALTinterwordstretchfactor}{4}
\providecommand{\BIBentryALTinterwordspacing}{\spaceskip=\fontdimen2\font plus
\BIBentryALTinterwordstretchfactor\fontdimen3\font minus \fontdimen4\font\relax}
\providecommand{\BIBforeignlanguage}[2]{{%
\expandafter\ifx\csname l@#1\endcsname\relax
\typeout{** WARNING: IEEEtran.bst: No hyphenation pattern has been}%
\typeout{** loaded for the language `#1'. Using the pattern for}%
\typeout{** the default language instead.}%
\else
\language=\csname l@#1\endcsname
\fi
#2}}
\providecommand{\BIBdecl}{\relax}
\BIBdecl

\bibitem{RISISAC_survey2}
M.~Rihan, A.~Zappone, and S.~Buzzi, ``Robust {RIS}-assisted {MIMO} communication-radar coexistence: Joint beamforming and waveform design,'' \emph{IEEE Trans. Commun.}, vol.~71, no.~11, pp. 6647--6661, 2023.

\bibitem{chen2}
Z.~Chen, G.~Chen, J.~Tang, S.~Zhang, D.~K. So, O.~A. Dobre, K.-K. Wong, and J.~Chambers, ``Reconfigurable intelligent-surface-assisted {B5G}/{6G} wireless communications: Challenges, solution, and future opportunities,'' \emph{IEEE Commun. Mag.}, vol.~61, no.~1, pp. 16--22, 2023.

\bibitem{czz}
Z.~Chen, J.~Tang, X.~Y. Zhang, D.~K.~C. So, S.~Jin, and K.-K. Wong, ``Hybrid evolutionary-based sparse channel estimation for {IRS}-assisted mmwave {MIMO} systems,'' \emph{IEEE Trans. Wireless Commun.}, vol.~21, no.~3, pp. 1586--1601, 2022.

\bibitem{RIS_ISAC_jiang}
Z.~Jiang, M.~Rihan, P.~Zhang, L.~Huang, Q.~Deng, J.~Zhang, and E.~M. Mohamed, ``{Intelligent reflecting surface aided dual-function radar and communication system},'' \emph{IEEE Syst. J.}, vol.~16, no.~1, pp. 475--486, 2022.

\bibitem{Zhongxin}
S.~Yan, S.~Cai, W.~Xia, J.~Zhang, and S.~Xia, ``{A reconfigurable intelligent surface aided dual-function radar and communication system},'' in \emph{IEEE Int. Symp. Jt. Commun. Sens., JC and S}, 2022, pp. 1--6.

\bibitem{RIS_ISAC_liu}
R.~Liu, M.~Li, Y.~Liu, Q.~Wu, and Q.~Liu, ``{Joint transmit waveform and passive beamforming design for RIS-aided DFRC systems},'' \emph{IEEE J. Sel. Top. Signal Process.}, vol.~16, no.~5, pp. 995--1010, 2022.

\bibitem{RIS_ISAC_fan}
X.~Meng, F.~Liu, S.~Lu, S.~P. Chepuri, and C.~Masouros, ``{RIS-assisted integrated sensing and communications: a subspace rotation approach: invited paper},'' in \emph{Proc. IEEE Radar. Conf.}, 2023, pp. 1--6.

\bibitem{RIS_supervised}
C.~Huang, G.~C. Alexandropoulos, C.~Yuen, and M.~Debbah, ``{Indoor signal focusing with deep learning designed reconfigurable intelligent surfaces},'' in \emph{Proc. IEEE 20th Int. Workshop Signal Process. Adv. Wireless Commun.}, 2019, pp. 1--5.

\bibitem{RIS_Un}
J.~Gao, C.~Zhong, X.~Chen, H.~Lin, and Z.~Zhang, ``Unsupervised learning for passive beamforming,'' \emph{IEEE Commun. Lett.}, vol.~24, no.~5, pp. 1052--1056, 2020.

\bibitem{RIS_Un_transfer}
Y.~Ge and J.~Fan, ``{Beamforming optimization for intelligent reflecting surface assisted MISO: a deep transfer learning approach},'' \emph{IEEE Trans. Veh. Technol.}, vol.~70, no.~4, pp. 3902--3907, 2021.

\bibitem{My}
Z.~Chen, J.~Ye, P.~Zhang, H.~Rizk, L.~Huang, and M.~Rihan, ``{A light-weight learning framework for RIS-assisted beamforming design with mobile edge computing},'' in \emph{IEEE/CIC Int. Conf. Commun. China}, 2023, pp. 1--6.

\bibitem{RL_MISO}
K.~Feng, Q.~Wang, X.~Li, and C.-K. Wen, ``Deep reinforcement learning based intelligent reflecting surface optimization for {MISO} communication systems,'' \emph{IEEE Wireless Commun. Lett.}, vol.~9, no.~5, pp. 745--749, 2020.

\bibitem{ISAC_predictbf}
J.~Mu, Y.~Gong, F.~Zhang, Y.~Cui, F.~Zheng, and X.~Jing, ``{Integrated sensing and communication-enabled predictive beamforming with deep learning in vehicular networks},'' \emph{IEEE Commun. Lett.}, vol.~25, no.~10, pp. 3301--3304, 2021.

\bibitem{ISAC_Un_bf}
X.~Liu, H.~Zhang, K.~Long, A.~Nallanathan, and V.~C.~M. Leung, ``{Distributed unsupervised learning for interference management in integrated sensing and communication systems},'' \emph{IEEE Trans. Wireless Commun.}, vol.~22, no.~12, pp. 9301--9312, 2023.

\bibitem{ISAC_RL}
A.~M. Ahmed, L.~Gharsalli, S.~Fortunati, and A.~Sezgin, ``{Reinforcement learning for cognitive integrated communication and sensing systems},'' in \emph{European Radar Conf.}, 2023, pp. 395--398.

\bibitem{closeform}
F.~Liu, Y.~Liu, A.~Li, C.~Masouros, and Y.~C. Eldar, ``{Cramér-Rao bound optimization for joint radar-communication beamforming},'' \emph{IEEE Trans. Signal Process.}, vol.~70, pp. 240--253, 2022.

\bibitem{mobilenet}
\BIBentryALTinterwordspacing
A.~G. Howard, M.~Zhu, B.~Chen, D.~Kalenichenko, W.~Wang, T.~Weyand, M.~Andreetto, and H.~Adam, ``Mobilenets: Efficient convolutional neural networks for mobile vision applications,'' \emph{arXiv:1704.04861}, 2017. [Online]. Available: \url{https://arxiv.org/abs/1704.04861}
\BIBentrySTDinterwordspacing

\end{thebibliography}
\end{document}